\documentstyle[aps,amsmath,amssymb,epsfig]{revtex}
\draft
\begin{document}
\tightenlines
\title{Microscopic Origin of Spatial Coherence and Wolf Shifts\footnote{Festschrift in honor of Prof. E. Wolf,
Edited by T. Jansen, SPIE Publication No (2004)}}
\author{Girish S. Agarwal\footnote{email: gsa@prl.ernet.in}}
\address{Physical Research Laboratory, Navrangpura, Ahmedabad-380 009, India}
\date{\today}
\maketitle
\begin{abstract}
\end{abstract}

\section{Introduction}
Wolf $^{1,2,3,4}$ discovered how the spatial coherence characteristics of the
source affect the spectrum of the radiation in the far zone. In
particular the spatial coherence of the source can result either
in red or blue shifts in the measured spectrum.His predictions
have been verified in a large number of different classes of
systems. Wolf and coworkers usually assume a given form of source
correlations and study its consequence. In this paper we consider
microscopic origin of spatial coherence and radiation from a
system of atoms$^{5,6,7,8}$. We discuss how the radiation is different from
that produced from an independent system of atoms. We show that
the process of radiation itself is responsible for the creation of
spatial correlations within the source. We present different
features of the spectrum and other statistical properties  of the
radiation, which show strong dependence on the spatial
correlations. We show the existence of a new type of two-photon
resonance that  arises as a result of such spatial correlations.
We further show how the spatial coherence of the field can be used
in the context of radiation generated by nonlinear optical
processes. We conclude by demonstrating the universality of Wolf
shifts and its application in the context of pulse propagation in
a dispersive medium.
\begin{figure}
\centerline{\psfig{figure=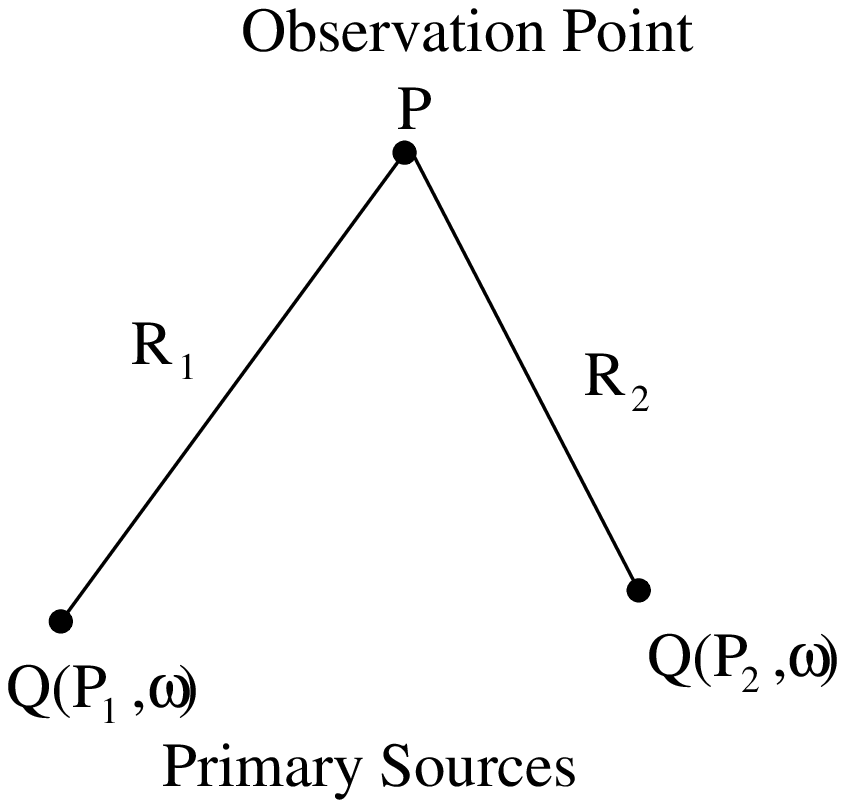,height=5cm,width=4cm}}
\caption{\label{fig1}}
\end{figure}
We start by giving a summary of Wolf's main results$^{1,2}$. Consider the radiation
produced by two point sources $P_1$ and $P_2$ at the observation point $P$,
 Fig.~\ref{fig1}. Let
us consider for simplicity the case of scalar fields $U(P,\omega)$. The spectrum
of the field at $P$ is given by
\begin{equation}
S_U(P,\omega)= \langle U^{*}(P,\omega)U(P,\omega)\rangle,
\label{1}
\end{equation}
where as the spectrum of the source is defined by
\begin{eqnarray}
S_{Q}(\omega)&=&\langle Q^{*}(P_1,\omega)Q(P_1,\omega)\rangle\\
&=&\langle Q^{*}(P_2,\omega)Q(P_2,\omega)\rangle.
\label{2}
\end{eqnarray}
We assume identical spectra for the two sources. Let $\mu_{Q}(\omega)$
be a measure of correlation between two sources:

\begin{eqnarray}
\mu_{Q}=\frac{\langle Q^{*}(P_1,\omega)Q(P_2,\omega)\rangle}{S_{Q}(\omega)},
\label{3}
\end{eqnarray}
The spectral degree of coherence between two sources would be $\mid\mu_{Q}\mid$.
For two coherent sources $\mu$ is $1$ whereas for incoherent sources $\mu=0$.
The field $U$ at the point $P$ can be related to the strength of the sources via
\begin{eqnarray}
U(P,\omega)=Q(P_1,\omega)\frac{e^{ikR_{1}}}{R_1}+
Q(P_2,\omega)\frac{e^{ikR_{2}}}{R_2}.
\label{4}
\end{eqnarray}
Here we have ignored unnecessary numerical factors. Using Eq.~(\ref{4}) the
spectrum of the field is related to the spectrum of the source and the degree of
spatial coherence
\begin{eqnarray}
S_U(P,\omega)=S_Q(\omega)\left(\frac{1}{R_1^2}+\frac{1}{R_2^2}+\frac{1}{R_1R_2}
\left[\mu_Q(\omega)e^{ik(R_{2}-R_1)}+{\rm c.
c.}\right]\right).
\label{5}
\end{eqnarray}
Clearly in general, the source spectrum and the spectrum at $P$ are not equal
\begin{equation}
S_U(P,\omega)\neq S_Q(\omega).
\label{6}
\end{equation}
Clearly the measured spectral characteristics will also be determined by
$\mu_{Q}$ and $S_U(P,\omega)$, in general, would exhibit correlation
induced spectral shifts. Wolf used phenomenological model for $S_Q$ and
$\mu_{Q}$ to demonstrate a variety of spectral shifts and even the
correlation induced splitting of a line into several lines. Clearly it is
desirable to understand the origin of source correlations.

\section{microscopic origin of source correlations} 

We thus examine the question
of how the atom radiate. Consider for example an atom in its excited state. It
interacts with the modes of quantized electromagnetic field in vacuum state. The
atom makes a transition to the ground state by the emission of a photon. The
photon can be emitted in any mode of the field. The atom has infinity of
available modes. It is known that the spectrum of the emitted radiation has
Lorentzian spectrum
\begin{equation}
S_A(\omega)=\frac{\gamma/\pi}{(\omega-\omega_0)^2+\gamma^2},
\label{7}
\end{equation}
where $\omega_0$ is the frequency of the atomic transition and
$\gamma$ is half the Einstein $A$ coefficient.
\begin{figure}
\centerline{\psfig{figure=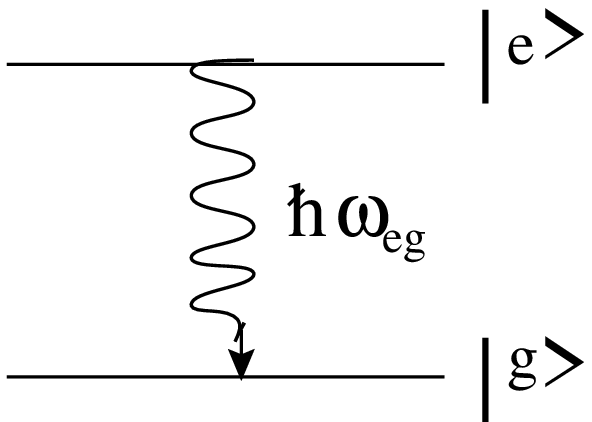,height=3cm,width=3cm}}
\caption{\label{fig2}}
\end{figure}
 Next consider two
atoms located at $\vec{r}_{A}$ and $\vec{r}_{B}$. Let each atom be
initially in its excited state. The question is whether the atoms
radiate independently of each other {\it i.e.} whether the
spectrum of the emitted photons factorizes
\begin{equation}
S(\omega_1,\omega_2)=S_A(\omega_1)S_B(\omega_2)
\label{8}
\end{equation}
or not. The correlations between the two atoms$^{6,7,8}$ would invalidate (\ref{8}) and
it would in general also imply that
\begin{eqnarray}
S_A(\omega_1)\neq\int S(\omega_1,\omega_2)d\omega_2,
\label{9}
\end{eqnarray}
\begin{figure}
\centerline{\psfig{figure=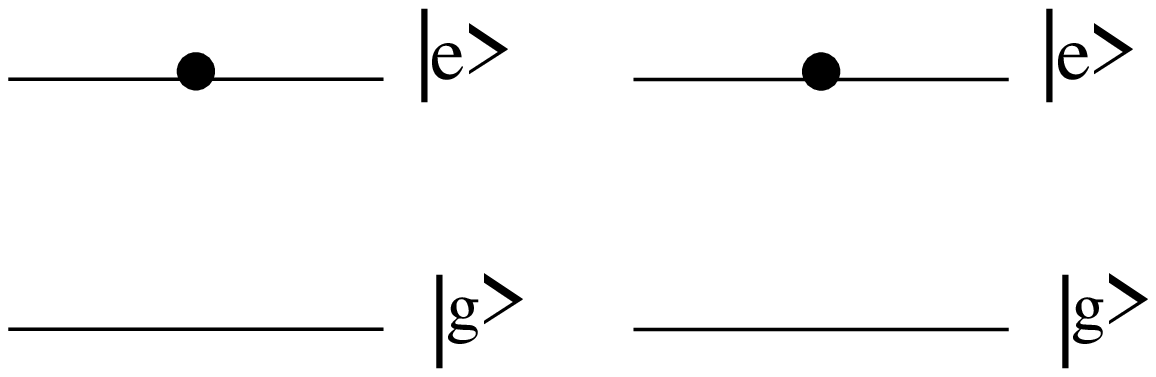,height=2cm,width=5cm}}
\caption{\label{fig3}}
\end{figure}
\noindent
{\it i.e.}, the spectrum of the emitted radiation would be different from the
one if the other atom was absent. Note that both atoms interact with a common
 quantized electromagnetic field. This interaction with a common field results
 in an effective interaction between two
 atoms even if the atoms do not interact. This can also be understood by
 considering, say, the net field on the atom $B(A)$ which would consist of the
 vacuum field and field radiated by the atom $A(B)$. Let us denote by
 $\chi_{ij}(\vec{r}_{A},\vec{r}_{B},\omega)$ as the $i-th$ component of the field at position
 $\vec{r}_{A}$ due to a unit dipole oriented in the direction $j$ at the position
 $\vec{r}_{B}$.
 \begin{figure}
\centerline{\psfig{figure=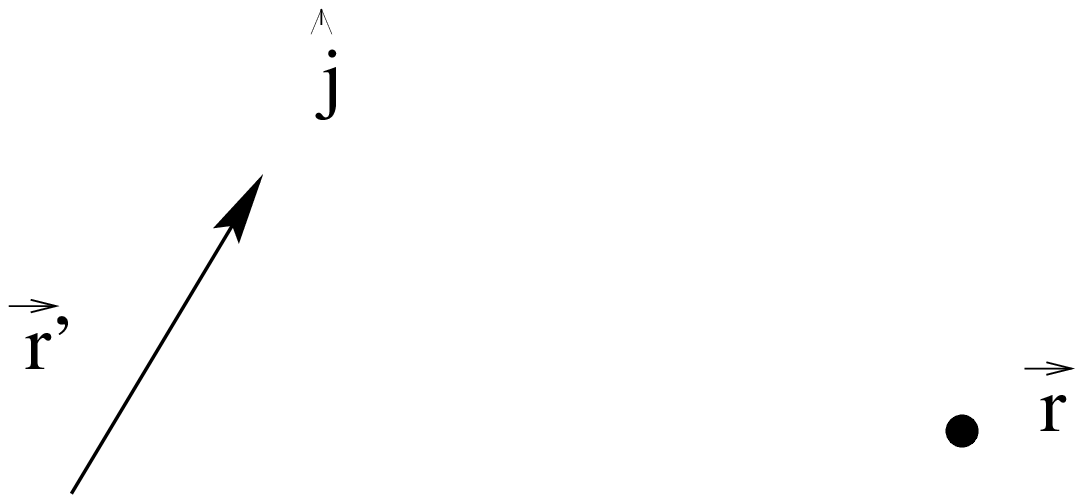,height=3cm,width=6cm}}
\caption{\label{fig4}}
\end{figure}
\noindent
 This field$^{9,10,11}$ is well known from the solution of Maxwell equations
 \begin{eqnarray}
 \chi_{ij}(\vec{r}_{A},\vec{r}_{B},\omega)=
 \left(\frac{\omega^2}{c^2}\delta_{ij}+\frac{\partial^2}
 {\partial r_{A}\partial r_{B}}\right)
 \frac{exp(i|\vec{r}_{A}-\vec{r}_{B}|\omega/c)}
 {|\vec{r}_{A}-\vec{r}_{B}|}.
 \label{10}
 \end{eqnarray}
 This function has close connection with the spatial coherence of the vacuum of
 the electromagnetic field. Let us write the electric field operator in terms of
 its positive and negative frequency parts
 \begin{eqnarray}
 E=E^{(+)}+E^{(-)}.
 \label{11}
 \end{eqnarray}
 It is well known in quantum optics that $E^{(+)}(E^{(-)})$
 corresponds to the
 absorption (emission) of photons. Further $E^{(+)}$ is an analytical signal.
 Let us consider second order coherence function of the electromagnetic field
 \begin{eqnarray}
 S^{A}_{\alpha\beta}(\vec{r_1},\vec{r_2},\tau)=\langle
 E_{\alpha}^{(+)}(\vec{r}_{1},t+\tau)E_{\beta}^{(-)}(\vec{r}_2)\rangle
 \label{12}
 \end{eqnarray}
 which is non-vanishing even-though the field is in vacuum state. Its Fourier
 transform is given by$^{9}$
 \begin{eqnarray}
 \int d\tau e^{i\omega\tau}S_{\alpha\beta}^{A}(\vec{r}_1,\vec{r}_2,\tau)&=&
 2\hbar Im\chi_{ij}(\vec{r_1},\vec{r_2},\omega){\rm~~~ if~~}\omega>0\nonumber\\
 &=&0{\rm~~~~if~~}\omega< 0.
 \label{13}
 \end{eqnarray}
 We thus conclude that the vacuum of the electromagnetic field has spatial
 coherence which extends over the dimensions of wavelength. Therefore the
 correlation between atoms would extend over at least distances of the order of
 wavelength. Clearly in a macroscopic sample these correlation could build up
 over much larger distances. Explicit results for two atoms can be found in
 Refs.~[6],[7],[8].
 \section{source correlation induced two photon resonance}
 \begin{figure}
\centerline{\psfig{figure=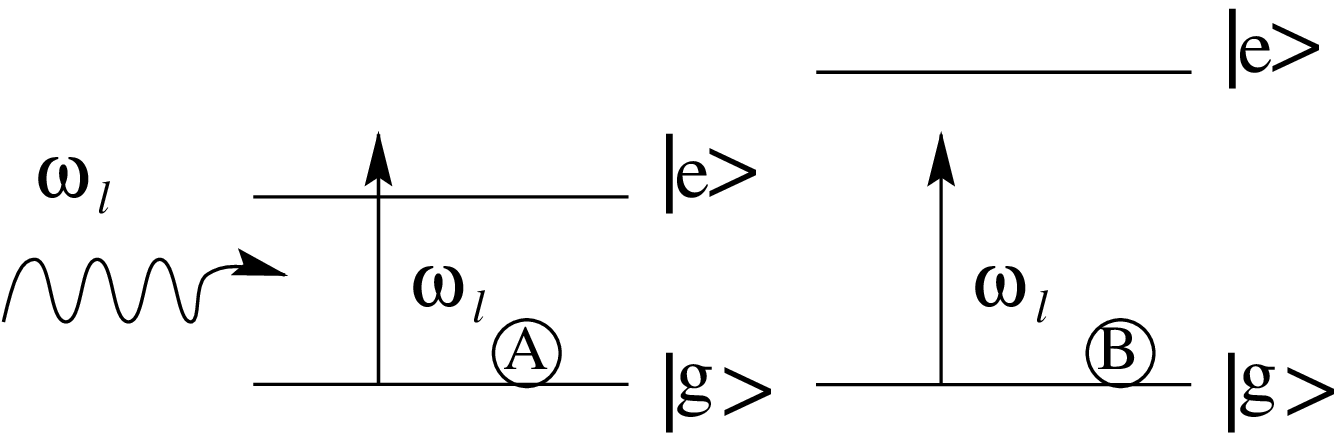,height=3cm,width=8cm}}
\caption{\label{fig5}}
\end{figure}
 We next discuss several other situations where atom-atom correlations play an
 important role. Consider first the case of two unidentical atoms with transition
 frequencies $\omega_A$ and $\omega_B$ and which are located within a wavelength
 of each other. Let both the atoms start in ground state and let these interact
 with a laser field of frequency $\omega_l$. We now study the total intensity
 $I(\omega_l)$ of the emitted radiation as a function of $\omega_l$. Clearly
 $I(\omega_l)$ will exhibit single photon resonance at
 $\omega_l=\omega_{Aeg},\omega_{Beg}$. In principle there is also the
 possibility of two photon resonance $2\omega_l=\omega_{Aeg}+\omega_{Beg}$. It
 turns out that in the absence of source correlations, the two photon resonance
 does not occur as the two paths
 \begin{eqnarray}
|g_A,g_B\rangle\rightarrow|e_A,g_B\rangle\rightarrow|e_A,e_B\rangle,{\rm~~and~~}
|g_A,g_B\rangle\rightarrow|g_A,e_B\rangle\rightarrow|e_A,e_B\rangle
\end{eqnarray}
interfere destructively. Thus the source correlations are the key to the two photon
 resonance. In an earlier work the effect of source correlations on such a two
 photon resonance was studied in great detail$^{7}$ and recently it has been observed
 in experiments involving single molecules$^{12}$ further very recently we show how the source
 correlation arise in a cavity$^{13}$.

 \section{spatial coherence and emission in presence of a mirror}
 \begin{figure}
\centerline{\psfig{figure=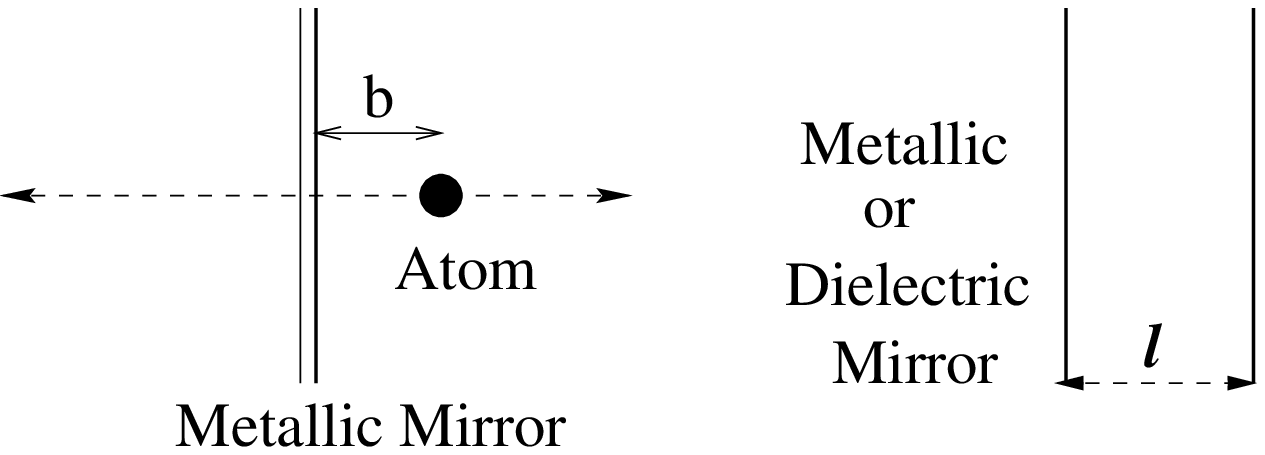,height=3cm,width=10cm}}
\caption{\label{fig6}}
\end{figure}
 Another class of systems where spatial coherence plays an important role is for
 example, the emission of radiation in front of a metallic mirror$^{10}$
 or in a cavity formed by metallic or dielectric mirrors. The spectrum
 of the emitted radiation depends on the distance of the atom from mirror. As a
 matter of fact both line width and line shift become $b$-dependent.
 If the metallic mirror is treated as a perfect conductor, then the calculations show
  that the line shifts, for example, are determined by the
 spatial coherence of the field at the location of the atom and its image. 
Thus the correlation of the vacuum
 $\langle\vec{E}^{(+)}(\vec{b},t)\vec{E}^{(-)}(-\vec{b},t^{'})\rangle$,
 which is related to $\chi(\vec{b},-\vec{b},\omega)$, determines the line shifts and line
 widths. Explicit results for the $b$ - dependence of shifts and widths  can be found in
 Refs.~[10],[14].

 \section{spatial coherence induced control of nonlinear generation}
 We next discuss the effects of spatial coherence in the context of nonlinear
 optics. We would show that the generation of radiation using nonlinear
 processes can be controlled by source correlations. Consider for example, the
 process of second harmonic generation (SHG) with $P=\chi^{(2)}E^2$, $E\sim
 e^{i\vec{k}.\vec{r}}$.
 The efficiency of the SHG depends on the phase matching integral
 \begin{eqnarray}
 f=\frac{1}{V}\int e^{-i\vec{q}.\vec{r}} e^{2i\vec{k}.\vec{r}}d^3r
 \label{14}
 \end{eqnarray}
 which goes to unity if $\vec{q}=2\vec{k}$.The function $f$ determines the direction in
 which second harmonic generation is dominant.
\begin{figure}
\centerline{\psfig{figure=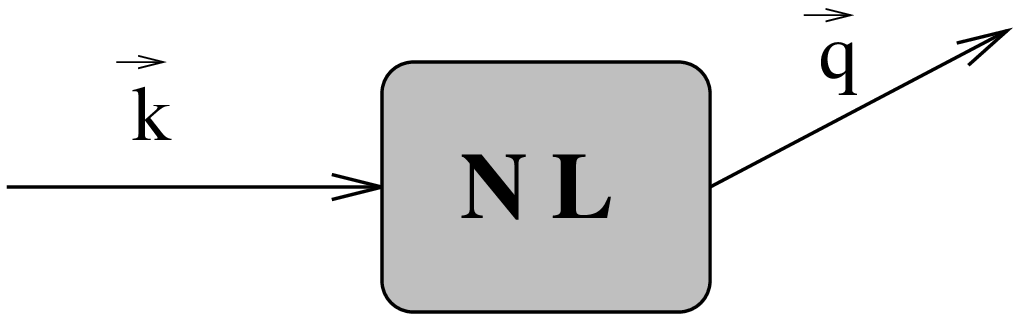,height=2cm,width=5cm}}
\caption{\label{fig7}}
\end{figure}
\noindent
 If however the field $E$ is
 partially coherent, then in place of (\ref{14}) we need to consider
 \begin{eqnarray}
 f=\int d^3r^{'}d^3r^{''}e^{-i\vec{q}.\vec{r}^{'}}
 e^{2i\vec{k}.\vec{r}^{''}}\langle
 P(\vec{r}^{'})P^{*}(\vec{r}^{''})\rangle.
 \label{15}
 \end{eqnarray}
  \begin{figure}
\centerline{\psfig{figure=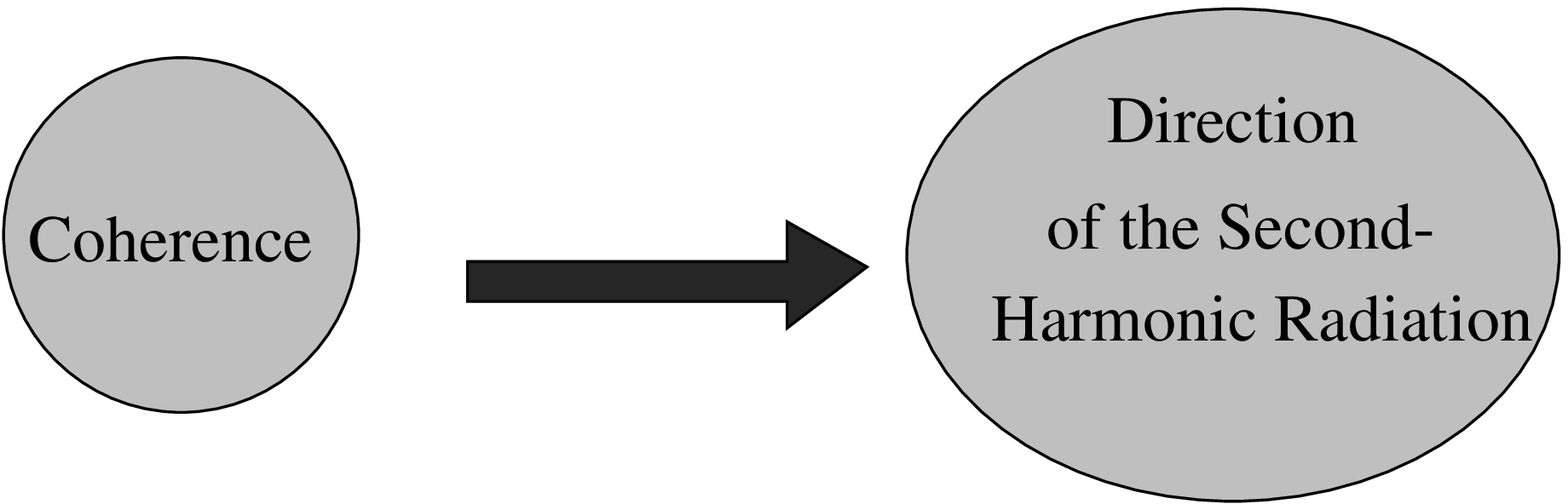,height=3cm,width=9cm}}
\caption{\label{fig8}}
\end{figure}
\noindent
 Note that for SHG with coherent radiation
 \begin{eqnarray}
 \langle P(\vec{r}^{'})P^{*}(\vec{r}^{''})\rangle\equiv
 \langle P(\vec{r}^{'})\rangle\langle
 P^{*}(\vec{r}^{''})\rangle\equiv e^{2i\vec{k}.(\vec{r}^{'}-\vec{r}^{''})}
 \label{16}
 \end{eqnarray}
 and then
 \begin{equation}
 I\propto (vol)^2.
 \end{equation}
 On the other hand for the case of incoherent radiation
 \begin{eqnarray}
 &&\langle P(\vec{r}^{'})P^{*}(\vec{r}^{''})\rangle\equiv|\wp|^2
 \delta(\vec{r}^{'}-\vec{r}^{''}),\\
 && I\rightarrow|\wp|^2(vol).
 \end{eqnarray}
 For the partially coherent radiation
 \begin{equation}
 \langle P(\vec{r}^{'})P^{*}(\vec{r}^{''})\rangle=|\chi^{(2)}|^2\langle
 E^2(\vec{r}^{'})E^{*2}(\vec{r}^{''})\rangle,
 \end{equation}
 which under the assumption of a Gaussian field will become
 \begin{eqnarray}
 \langle P(\vec{r}^{'})P(\vec{r}^{''})\rangle=
 2I^2|\mu(\vec{r}^{'}-\vec{r}^{''})|^2,
 \label{21}
 \end{eqnarray}
 where $\mu(\vec{r}^{'}-\vec{r}^{''})$ denotes the degree of spatial coherence
 of the incident field.
 Thus SHG would now be determined by the integral
 \begin{eqnarray}
 &&|f(\vec{Q})|^2=\int\int
 d^3r^{'}d^3r{''}|\mu(\vec{r}^{'}-\vec{r}^{''})|^2
 e^{\vec{Q}.(\vec{r}^{'}-\vec{r}^{''})}\\
 &&\vec{Q}=-\vec{q}+2\vec{k}.
 \label{22}
 \end{eqnarray}
 Clearly now the direction of SHG would be determined by the spatial coherence
 of the field. Thus spatial coherence can serve as a control parameter for the
 nonlinear generation. Clearly the above ideas should also find interesting
 applications in other areas of nonlinear optics as well.

\section{universality of Wolf shift}
Before concluding the paper we also like to make some general
remarks for the universality and applicability of Wolf shifts in
the context of other systems. We know for instance, that other
standard equations of physics (such as those describing vibrations
of string, heat transport) admit following relation between the
effect $\Phi$ of the source $P$ at the observation point
\begin{eqnarray}
\Phi(\vec{r})=\int G(\vec{r},\vec{r}^{'})P(\vec{r}^{'})d^3r^{'},
\label{23}
\end{eqnarray}
where $G$ is Green's function for the underlying equation. The observed
quantities are usually quadratic in $\Phi$. Thus observation at the point
$\vec{r}$ would depend on the correlations of the source at two points. This is
due to the nonlocal nature of the solution (\ref{23}).

\section{Fluctuating Pulses in a Dispersive medium}
As another example of this universality we can consider the propagation of
pulses in a dispersive medium which is described by the equation
\begin{eqnarray}
i\frac{\partial{\it E}}{\partial z}=\frac{\tilde{k}}{2}\frac{\partial^{2}
{\it E}}{\partial t^2}
\label{24}
\end{eqnarray}
The solution of this equation can be given in terms of Green's function
\begin{eqnarray}
&&{\it E}(z,t)=\int G(z,t;0,t^{'}){\it E}(0,t^{'})dt^{'};\\
&&G=\frac{i}{2\pi z\tilde{k}}exp\left(-
\frac{i}{2z\tilde{k}}(t^2-2tt^{'}+t^{'2})\right).
\label{25}
\end{eqnarray}
If the input pulse has fluctuations, then the intensity of the output pulse
would be determined by the correlation in pulses on input plane
\begin{eqnarray}
I(L,t)=\int\int dt{'}dt{''}G^{*}(L,t;0,t^{'})G(L,t;0,t^{''})
\langle{\it E}(t{'}){\it E}^{*}(t{''})\rangle.
\label{26}
\end{eqnarray}
Clearly the intensity of the pulse at the output plane is not completely determined by the
intensity of the pulse at input plane.

\section{conclusions}
Thus in conclusion we have shown that the vacuum of
electromagnetic field has intrinsic partial spatial coherence in
frequency domain which effectively extends over regions of the
order of wavelength $\lambda$. This spatial coherence leads to a
dynamical coupling between atoms and is the cause of source
correlations. We showed how such correlations can lead to a new
type of two photon resonance and how these are relevant for near
field optics. We further showed how the source spatial
correlations can lead to new phase matching conditions for
nonlinear optical effects leading to the possibility of using
spatial coherence to produce tailor made emissions. We also
discussed the universality of source correlation effects and as a
specific example we treated the case of the propagation of
fluctuating pulses in a dispersive medium.

The author thanks E. Wolf for many discussions on the subject of
correlation induced shifts.


\begin{references}
\bibitem{1}
E. Wolf, "Invariance of the Spectrum of Light on Propagation," {\it Phys. Rev. Lett.\/} {\bf 56}, pp. 1370-1372 (1986);
 E. Wolf, "Red shifts and blue shifts of spectral lines emitted by two correlated sources," {\it ibid.\/} {\bf 58}, pp. 2646-2648 (1987); 
E. Wolf, "Correlation-induced Doppler-type frequency shifts of spectral lines," {\it ibid.\/} {\bf 63}, pp. 2220-2223 (1989).

\bibitem{2} 
E. Wolf, "Non-cosmological redshifts of spectral lines," {\it Nature (London)} {\bf 326}, pp. 363-366 (1987).
\bibitem{3}
E. Wolf and D. F. V. James, "Correlation-induced spectral changes," {\it Rep. Prog. Phys.} {\bf 59}, pp. 771-818 (1996).
\bibitem{4} L. Mandel and E. Wolf, {\it Optical Coherence and Quantum Optics} (Cambridge University Press, 1995).
\bibitem{5} G. S. Agarwal, in {\it Quantum Optics} (Springer Tracts in Modern
Physics, Vol. 70, 1974).
\bibitem{6} G. Varada and G. S. Agarwal, "Microscopic approach to correlation-induced frequency shifts," {\it Phys. Rev. A} {\bf 44}, pp. 7626-7634 (1991).
\bibitem{7} G. Varada and G. S. Agarwal, "Two-photon resonance induced by the dipole-dipole interaction," {\it Phys. Rev. A} {\bf 45}, pp. 6721-6729 (1992).
\bibitem{8} D. F. V. James, "Frequency shifts in spontaneous emission from two interacting emission," {\it Phys. Rev. A} {\bf 47}, pp. 1336-1346 (1993).
\bibitem{9} G. S. Agarwal, "Quantum electrodynamics in the presence of dielectrics and conductors : I. Electromagnetic-field response functions and black-body fluctuations in finite geometries," {\it Phys. Rev. A} {\bf 11}, pp. 230-242 (1975).
\bibitem{10} G. S. Agarwal, "Quantum electrodynamics in the presence of dielectrics and conductors : IV. General Theory of spontaneous emission in finite geometries," {\it Phys. Rev. A} {\bf 12}, pp. 1475-1497 (1975).
\bibitem{11} G. S. Agarwal, in {\it Quantum Electrodynamics and Quantum Optics},
ed. A. Barut (Plenum, 1983).
\bibitem{12}C. Hettich, C. Schmitt, J. Zitzmann, S. Kuhn, I. Gerhardt, and
V. Sandoghdar, "Nanometer resolution and and coherent optical dipole coupling of two individual molecules," {\it  Science} {\bf 298}, pp. 385-389 (2002).
\bibitem{13} P. K. Pathak and G. S. Agarwal, "Giant two-atom two-photon vacuum Rabi oscillations in a high quality cavity," to be published.
\bibitem{14} G. S. Agarwal and H. D. Vollmer, "Surface polariton effects in spontaneous emission," {\it Physica Status Solidi B}
{\bf 79}, 249 (1977); G. S. Agarwal and Vollmer, "Surface polariton effects in spontaneous emission. II. Effects of spatial dispersion," {\it ibid.\/} {\bf 85}, 301
(1978).
\end{references}
\end{document}